\documentclass[a4paper,11pt]{article}

\usepackage{jcappub} 

\usepackage[T1]{fontenc} 

\usepackage{color, colortbl}
\definecolor{LightCyan}{rgb}{0.88,1,1}
\definecolor{Gray}{gray}{0.9}

\usepackage[utf8]{inputenc}

\title{\boldmath A low Hubble Constant from  galaxy distribution observations}

\author[a]{R. F. L. Holanda}
\author[a,b,1]{G. Pordeus-da-Silva,\note{Corresponding author.}}
\author[c]{and S. H. Pereira}

\affiliation[a]{Departamento de F\'isica Te\'orica e Experimental - Universidade
	Federal do Rio Grande do Norte, Natal - RN, Brasil.}
	
\affiliation[b]{ Escola CAIC José Joffily, Secretaria da Educação e da Ciência e Tecnologia, Governo da Paraíba, Campina Grande - PB, Brasil.}

\affiliation[c]{Departamento de F\'isica e Qu\'imica, Universidade Estadual Paulista (UNESP), Faculdade de Engenharia de Guaratinguet\'a, Av. Ariberto Pereira da Cunha, 333 -  Guaratinguet\'a - SP, Brasil.}

\emailAdd{holandarfl@fisica.ufrn.br}
\emailAdd{givalpordeus@ufrn.edu.br}
\emailAdd{s.pereira@unesp.br}

\abstract{An accurate determination of the Hubble constant remains a puzzle in observational cosmology. The possibility of a new physics has emerged with a significant tension  between the current expansion rate of our Universe measured from the cosmic microwave background by the Planck satellite and from local methods. In this paper, new tight estimates on this parameter are obtained by considering two data sets from galaxy distribution observations: galaxy cluster gas mass fractions and baryon acoustic oscillation measurements. Priors from the Big Bang nucleosynthesis (BBN) were also considered. By considering the flat  $\Lambda$CDM and XCDM models,  and the non-flat $\Lambda$CDM model, our main results are: $H_0=65.9^{+1.5}_{-1.5}$ km s$^{-1}$ Mpc$^{-1}$, $H_0=65.9^{+4.4}_{-4.0}$ km s$^{-1}$ Mpc$^{-1}$ and $H_0=64.3^{+ 4.5}_{- 4.4}$ km s$^{-1}$ Mpc$^{-1}$  in $2\sigma$ c.l., respectively. These estimates are in full agreement  with the Planck satellite results. Our analyses in these cosmological scenarios also support a negative value for the deceleration parameter at least in 3$\sigma$ c.l..
}

\keywords{Hubble Constant, Galaxy Distribution, Cosmological Parameters}

\arxivnumber{2006.06712}

\begin{document}

\maketitle

\flushbottom

\section{Introduction}

During the past decades, the efforts of observational cosmology have been mainly focused on a precise determination of the parameters that describe the evolution of the Universe. Undoubtedly, one of the most important quantities to understand the cosmic history is the current expansion rate $H_0$, which is fundamental to answer important questions concerning different phases of cosmic evolution, as a precise determination of the cosmic densities, the mechanism behind the primordial inflation as well as the current cosmic acceleration~(see \cite{suyu2012hubble} for a broad discussion). 

Nowadays, the most reliable measurements of the Hubble constant are obtained from distance measurements of galaxies in the local Universe using Cepheid variables and Type Ia Supernovae (SNe Ia), which furnishes $H_0=74.03\pm 1.42$ km s$^{-1}$ Mpc$^{-1}$ \citep{Riess:2019cxk}. The value of $H_0$ can also be estimated from a cosmological model fit to the cosmic microwave background (CMB) radiation anisotropies. By assuming the flat $\Lambda$CDM model, the $H_0$ estimate is $H_0=67.36\pm 0.54$ km s$^{-1}$ Mpc$^{-1}$ \citep{Aghanim:2018eyx}\footnote{Another recent estimate of $H_0$ has been reported by the H0LiCOW collaboration \citep{rusu2019h0licow} based on lensing time-delays observations, $H_0 = 71.9^{+2.4}_{-3.0}$ km s$^{-1}$ Mpc$^{-1}$, which is in moderate tension with Planck. However, when combined with clustering data, a value of $H_0 = 66.98 \pm 1.18$ km s$^{-1}$ Mpc$^{-1}$ is obtained.}. These two $H_0$ values are discrepant by $\simeq 4.4\sigma$, which gives rise to the so-called $H_0$-tension problem\footnote{We recommend \cite{Freedman:2017yms} for an overview and history, as well as \cite{Verde_2019} for the current state of this intriguing problem.}. 

For this reason, new models beyond the standard cosmological one (the flat $\Lambda$) that could alleviate this tension become appealing. Some extensions of the $\Lambda$CDM model that allow to reduce the $H_0$ tension are: the existence of a new relativistic particle \citep{D_Eramo_2018}, small spatial curvature effects \citep{Bolejko_2018}, evolving dark energy models \citep{M_rtsell_2018}, among others \citep{Riess:2019cxk}. Then, new methods to estimate $H_0$ are welcome in order to bring some light on this puzzle. Precise measurements of the cosmic expansion rate $H(z)$ are important to provide more  restrictive constraints on cosmological parameters as well as  new insights into some fundamental questions that range from the mechanism behind the primordial inflation and current cosmic acceleration to neutrino physics~(see. e.g., \citep{suyu2012hubble} for a broad discussion).

The Hubble constant has also been estimated from galaxy cluster systems by using their angular diameter distances obtained from the Sunyaev-Zel'dovich effect (SZE) plus X-ray observations. For instance, \cite{Reese_2002} used 18  angular diameter distances of galaxy clusters  with redshifts ranging from $z =0.14$ up to $z=0.78$ and obtained $H_0= 60 \pm 4$ km s$^{-1}$ Mpc$^{-1}$ (only statistical errors) for an $\{\Omega_\text{m}=0.3$, $\Omega_{\Lambda}=0.7\}$ cosmology. The authors of the Ref.~\cite{Bonamente_2006} considered  38  angular diameter distances of galaxy clusters  in the redshift range $0.14 \leq z \leq 0.89$ and obtained $H_0= 76.9 \pm 4$ km s$^{-1}$ Mpc$^{-1}$ (only statistical errors) also for an $\{\Omega_\text{m}=0.3$, $\Omega_{\Lambda}=0.7\}$ cosmology. In both cases, it was assumed a spherical morphology to describe the clusters. Without fixing cosmological parameters, the authors of the Ref.~\cite{Holanda_2012} estimated $H_0$ by using a sample of angular diameter distances of 25 galaxy clusters (described by an elliptical density profile) jointly with baryon acoustic oscillations (BAO) and the CMB Shift Parameter signature. {The $H_0$ value obtained in the framework of $\Lambda$CDM model with arbitrary curvature was $H_0 = 74^{+8.0}_{-7.0}$  km s$^{-1}$ Mpc$^{-1}$ at 2$\sigma$ c.l.. By considering a flat $w$CDM model with a constant equation of state parameter, they obtained $H_0 = 72^{+10}_{-9.0}$ km s$^{-1}$ Mpc$^{-1}$ at 2$\sigma$ c.l.. In both cases were considered the statistical and systematic errors. As one may see, due to large error bars, the results found are in agreement with the current Riess et al. local estimate \citep{Riess:2019cxk} and with the Planck satellite estimate within 2$\sigma$.} It is worth to comment that the constraints on the Hubble constant via X-ray surface brightness and SZE observations of the galaxy clusters depend on the validity of the cosmic distance duality relation (CDDR): $D_L(1+z)^{-2}/D_A=1$, where $D_L$ is the luminosity distance and $D_A$ is the angular diameter distance \citep{Uzan_2004, HOLANDA_2012b}.

In this paper, we obtain new and tight estimates on the Hubble constant by combining  two main data sets from galaxy distribution observations in redshifts: 40 cluster galaxy gas mass fractions (GMF) and 11 baryon acoustic oscillation (BAO) measurements. Priors from BBN to calibrate the cosmic sound horizon and the  cosmic microwave background local  temperature as given by the FIRES/COBE also are considered. The $H_0$ estimates are performed in three models: flat $\Lambda$CDM and XCDM models,  and non-flat $\Lambda$CDM model. This last one is motivated by recent discussions in the literature concerning a possible cosmological curvature tension, with the Planck CMB spectra preferring a positive curvature at more than 99\% c.l. \citep{Di_Valentino_2019,George,Di_Valentino2}.  We show that the combination of these two independent data sets provides an interesting method to constrain the Hubble constant. For all models, tight estimates are found and our results support low Hubble constant values in agreement with the Planck results. We also found that the analyses performed in these cosmological models  indicate an universe in accelerated expansion in more than $3\sigma$ c.l.. 

The paper is organized as follows: Section~\ref{Section2} presents the two cosmological models and data sets used. Section~\ref{Section3} presents the main results and analysis and Section~\ref{Section4} finishes with conclusions.

\section{Cosmological Models and data sets.}\label{Section2}

In order to estimate the Hubble constant, we consider three cosmological scenarios: the flat  $\Lambda$CDM and XCDM models, and the non-flat $\Lambda$CDM model. Both $\Lambda$CDM models consider the cosmic dynamics dominated by a cold dark matter (CDM) component and cosmological constant ($\Lambda$), usually related to the constant vacuum energy density with negative pressure. By considering a constant equation of state for dark energy, $p_{\Lambda}=-\rho_{\Lambda}$, and the Universe described by a homogeneous and isotropic Friedmann-Lemaître-Robertson-Walker geometry, we obtain  from the Einstein equation the following expression for the Hubble parameter in the $\Lambda$CDM framework :
\begin{equation}
H(z)=H_{0}\sqrt{\Omega_{\text{m,}0}(1+z)^{3}+\Omega_{\text{k,}0}(1+z)^{2}+\Omega_\Lambda} \,, 
\label{Hz}
\end{equation}
where $H_0$ is the current Hubble constant, generally expressed in terms of the dimensionless parameter $h\equiv H_{0}/(100$ km s$^{-1}$ Mpc$^{-1})$, $\Omega_{\text{m,}0}$, $\Omega_\Lambda$ and $\Omega_{\text{k,}0}$ are the current dimensionless parameter of matter density (baryons + dark matter), dark energy density  and curvature density ($\Omega_{\text{k,0}}\equiv 1-\Omega_{\text{m,}0}-\Omega_\Lambda$), respectively. Note that if  $\Omega_{\text{k,0}}=0$ the flat $\Lambda$CDM model is recovered.

By considering the flat XCDM model, the Hubble parameter is written as: 
\begin{equation}
H(z)=H_{0}\sqrt{\Omega_{\text{m,}0}(1+z)^{3}+(1-\Omega_{\text{m,}0})^{3(1+\omega_x)}} \,,
\label{Hzx}
\end{equation}
where $\omega_x$ is the  equation of state parameter of dark energy, considered as a constant in our work (if $\omega_x=-1$, the flat $\Lambda$CDM model is recovered).

\subsection{Data sets and $\chi^{2}$ function}

In this section, we present the data sets used in the statistical analyses and their respective $\chi^{2}$ function.

\subsubsection{Sample I: Baryonic Acoustic Oscillation data}

This sample is composed of 11 measures obtained by 7 different surveys presented in Table \ref{tabBAO}. The relevant physical quantities for the BAO data are the angular diameter distance\footnote{The Robustness of baryon acoustic oscillations constraints in models beyond the flat $\Lambda$CDM model have been verified and discussed in details, for instance, in the works of \cite{bao1} and \cite{bao2}.}:
\begin{equation}
D_{\text{A}}(z) = \frac{1}{(1+z)}\times 
\left\{ \begin{array}{lll} \frac{H_{0}^{-1}}{\sqrt{|-\Omega_{\text{k,0}}|}}\text{sin}\left(\frac{\sqrt{|-\Omega_{\text{k,0}}|}}{H_{0}^{-1}}\int_{0}^{z}\frac{dz'}{H(z')} \right) & \text{if} &  \Omega_{\text{k,0}}<0  \\
									\int_{0}^{z}\frac{dz'}{H(z')} & \text{if} &  \Omega_{\text{k,0}}= 0  \\
                                    \frac{H_{0}^{-1}}{\sqrt{|-\Omega_{\text{k,0}}|}}\text{sinh}\left(\frac{\sqrt{|-\Omega_{\text{k,0}}|}}{H_{0}^{-1}}\int_{0}^{z}\frac{dz'}{H(z')} \right) & \text{if} &  \Omega_{\text{k,0}}>0  
\end{array} \right . \ ,
\label{DA z}
\end{equation}
the spherically-averaged distance:
\begin{equation}
    D_{\text{V}}(z)=\left[(1+z)^{2}D^{2}_{\text{A}}(z)\frac{z}{H(z)} \right]^{1/3}
\end{equation}
and the sound horizon at the drag epoch \citep{Eisenstein_1998}:
\begin{equation}
    r_{\text{s}}(z_{\text{d}})=\frac{2}{3k_{\text{eq}}}\sqrt{\frac{6}{R(z_{\text{eq}})}}\text{ln}\left[\frac{\sqrt{1+R(z_{\text{d}})}+\sqrt{R(z_{\text{d}})+R(z_{\text{d}})}}{1+\sqrt{R(z_{\text{eq}})}}\right] ,
\end{equation}
where $z_{\text{d}}$ is the drag epoch redshift, $z_{\text{eq}}$ is the equality redshift, $k_{\text{eq}}$ is the scale of the particle horizon at the equality epoch and \citep{Eisenstein_1998}
\begin{equation}
    R(z)\equiv \frac{3\rho_{\text{b}}}{4\rho_{\gamma}}=31.5(\Omega_{\text{b}}h^{2})\left(\frac{T_{\text{CMB}}}{2.7 \ \text{K}}\right)^{-4}\left(\frac{z}{10^3}\right)^{-1}
\end{equation} 
is the ratio of the baryon to photon momentum density. Here we determine $z_{\text{eq}}$, $k_{\text{eq}}$ and $z_{\text{d}}$, using the fit obtained by \cite{Eisenstein_1998}. For the CMB temperature, we use the value measured by COBE / FIRAS \citep{Fixsen_1996}, that is, $T_{\text{CMB}}=2.725$ K. This value is independent of the Planck satellite analyses  \cite{Addison_2018}.

\begin{table}
\caption{BAO data set consisting of 1 measurement of the survey 6dFGS \citep{Beutler:2011hx}, 1 of SDSS-LRG \citep{Padmanabhan:2012hf}, 1 of BOSS-MGS \citep{Ross:2014qpa}, 1 of BOSS-LOWZ \citep{Anderson:2014}, 1 of BOSS-CMASS \citep{Anderson:2014}, 3 of BOSS-DR12 \citep{Alam:2017hwk} and 3 of WiggleZ \citep{Kazin:2014qga}. The BAO variable $\mathcal{D}(z)$, $\sigma_{\mathcal{D}}$ and $r_s^{\rm fid}$ have units of Mpc, while $d_z(z)$ and $\sigma_{d_z}$ are dimensionless.}
\centering
\begin{tabular}{cccccc}
\hline \hline
Survey Set I & $z$ &  $d_z (z)$ & $\sigma_{d_z}$ & --  \\ \hline
 6dFGS & 0.106 & 0.336 & 0.015& --     \\
SDSS-LRG  & 0.35 & 0.1126 & 0.0022 & --  \\ \hline
 & & & & \\
\hline
Survey Set II & $z$ & $ \mathcal{D}(z)$ & $\sigma_{\mathcal{D}}$ & $r_s^{\rm fid}$  \\ \hline 
 BOSS-MGS  & 0.15 & 664 & 25 & 148.69 \\ 
BOSS-LOWZ  & 0.32 & 1264 & 25 & 149.28 \\ 
 BOSS-CMASS  & 0.57 & 2056 & 20 & 149.28 \\ 
 BOSS-DR12  & 0.38 & 1477 & 16 & 147.78 \\ 
  BOSS-DR12  & 0.51 & 1877 & 19 & 147.78 \\
 BOSS-DR12 & 0.61 & 2140 & 22 & 147.78 \\ 
  WiggleZ  & 0.44 & 1716 & 83 & 148.6 \\
 WiggleZ  & 0.60 & 2221 & 101 & 148.6 \\
  WiggleZ  & 0.73 & 2516 & 86 & 148.6 \\ \hline \hline
\end{tabular}
\label{tabBAO}
\end{table}

For survey set I, the BAO quantity is given by:
\begin{equation}
d_z(z)=\frac{r_{\text{s}}(z_{\text{d}})}{D_{\text{V}}(z)} \,, \label{dz}
\end{equation}
with a $\chi^{2}$ function given by:
\begin{equation}
\chi^{2}_{\text{BAO,I}}=\sum_{i=1}^{2} \left[\frac{d_{z}^{\text{th}}(z_{i})-d_{z, \ i}^{\text{ob}}}{\sigma_{d_{z , \ i}^{\text{ob}}}}\right]^2 \,.   
\end{equation}

On the other hand, the BAO quantity for survey set II is given by:
\begin{equation}
\mathcal{D}(z)=\frac{D_{\text{V}}(z)}{r_{\text{s}}(z_{\text{d}})} r_s^{\rm fid} \,, \label{alpha}
\end{equation}
and, in this specific case, the $\chi^{2}$ function is:
\begin{align}
\chi^{2}_{\text{BAO,II}} = &\sum_{i=1}^{6} \left[\frac{\mathcal{D}^{\text{th}}(z_{i})-\mathcal{D}_{i}^{\text{ob}}}{\sigma_{\mathcal{D}_{i}^{\text{ob}}}}\right]^2  \nonumber \\
&  + \left[\vec{\mathcal{D}}^{\text{th}}-\vec{\mathcal{D}}^{\text{ob}}\right]^{\text{T}}\text{\textbf{C}}^{-1}_{\text{WiggleZ}}\left[\vec{\mathcal{D}}^{\text{th}}-\vec{\mathcal{D}}^{\text{ob}}\right] \,,   
\end{align}
where $\text{\textbf{C}}^{-1}_{\text{WiggleZ}}$ is the inverse covariance matrix, whose explicit form is \citep{Kazin:2014qga}:
\begin{equation}
10^{-4}\left( 
\begin{array}{rrr}  2.17898878  & -1.11633321    & 0.46982851 \\ 
                    -1.11633321  & 1.70712004    & -0.71847155 \\
                    0.46982851   & -0.71847155   & 1.65283175 \end{array} \right).
\end{equation}
Unlike the others, the data points of the WiggleZ survey are correlated. 

\subsubsection{Sample II: Galaxy Cluster Gas Mass Fractions}

\begin{figure}
\centering
	\includegraphics[scale=0.55]{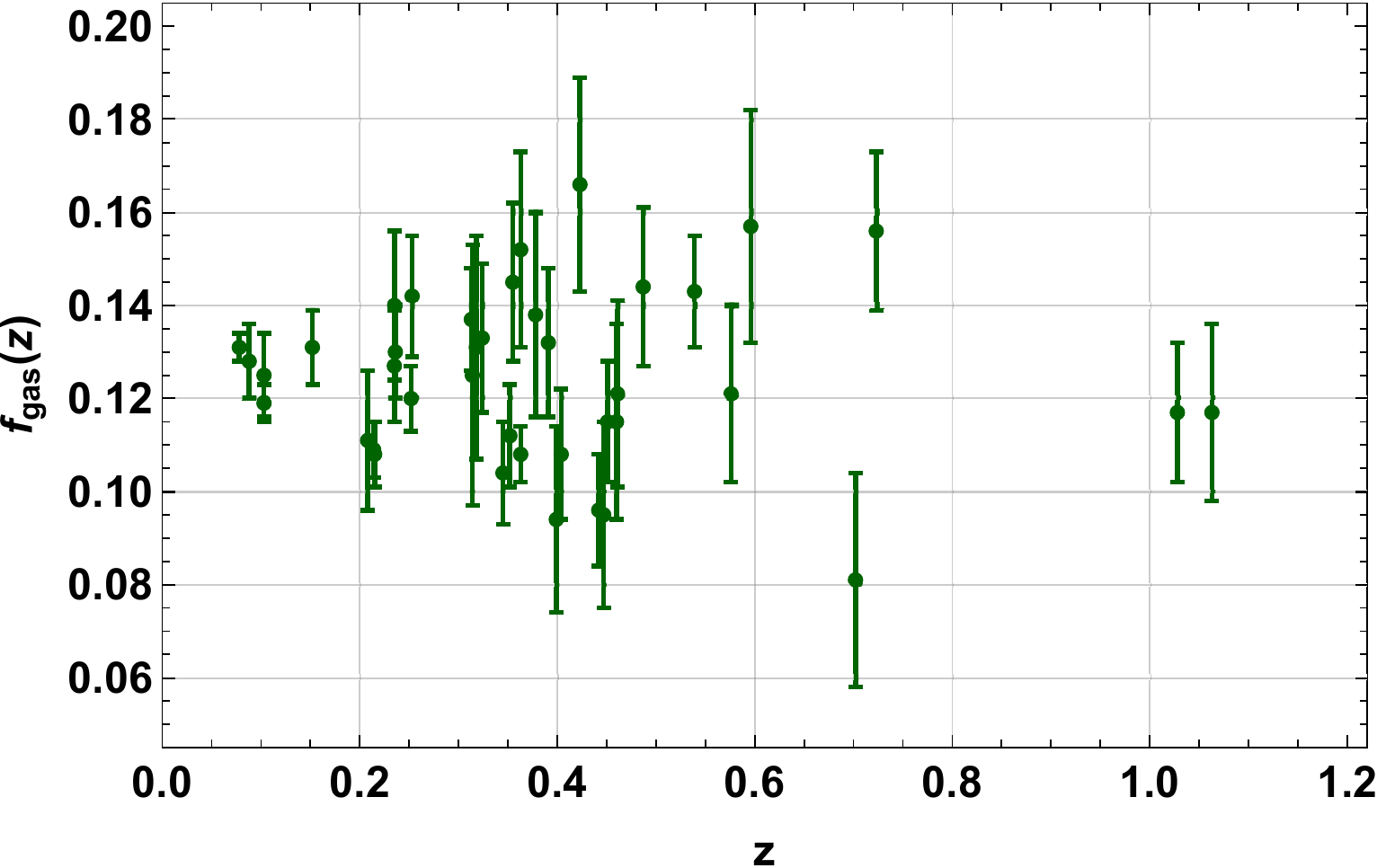}
    \caption{Measurements of $f_{\text{gas}}\left( z\right)$ used in our analysis. Details on this sample are presented in Table~2 of the Ref.~\cite{Mantz:2014xba}.}
    \label{fig:GMF}
\end{figure}

The  gas mass fractions (GMF) considered in this work corresponds to 40 Chandra observations  from massive and dynamically relaxed galaxy clusters in redshift range $0.078 \leq z \leq 1.063$ from the Ref.~\cite{Mantz:2014xba} (see Figure~\ref{fig:GMF}). These authors incorporated a robust gravitational lensing calibration of the X-ray mass estimates. The measurements of the gas mass fractions were performed in spherical shells at radii near $r_{2500}$\footnote{This radii is that one within which the mean cluster density is 2500 times the critical density of the Universe at the cluster's redshift.}, rather than  integrated at all radii ($< r_{2500}$). This approach significantly reduces systematic uncertainties compared to previous works that also estimated galaxy cluster gas mass fractions. 

The gas mass fraction quantity for a cluster is given by \citep{Mantz:2014xba,Allen_2011}:
\begin{equation}
f_{\text{gas}}^{\text{X-ray}}\left( z\right) =A(z)K(z)\gamma (z) \frac{\Omega _{\text{b}}(z)}{\Omega _{\text{m}}(z)} \left[ \frac{D_{A}^{\text{fid}}\left( z\right) }{D_{A}\left( z\right) }\right] ^{\frac{3}{2}} \ , \label{Eq fgas}
\end{equation} 
where
\begin{equation}
    A(z)=\left[\frac{H(z)D_{A}(z)}{H^{\text{fid}}(z)D^{\text{fid}}_{A}(z)} \right]^{\eta}
\end{equation}
stands for the angular correction factor ($\eta=0.442\pm 0.035$), $\Omega_{\rm{m}}(z)$ is the total mass density parameter, which corresponds to the sum of the baryonic mass density parameter, $\Omega_{\rm{b}}(z)$, and the dark matter density parameter, $\Omega_{\rm{c}}(z)$. The term in brackets corrects the angular diameter distance $D_A(z)$ from the fiducial model used in the observations, $D_A^\mathrm{fid}(z)$, which makes these measurements model-independent. The parameters $\gamma(z)$ and $K(z)$  correspond, respectively, to the depletion factor, i.e., the rate by which the hot gas fraction measured in a galaxy cluster is depleted with respect to the baryon fraction universal mean and to the bias of X-ray hydrostatic masses due to both astrophysical and instrumental sources. We adopt the value of $\gamma=0.848 \pm 0.085$ in our analysis, which was obtained from hydrodynamical simulations \citep{Planelles_2013} (see also a detailed discussion in section~4.2 in the Ref.~\cite{Mantz:2014xba}). The $\gamma$ parameter has also been estimated via observational data (SNe Ia, gas mass fraction, Hubble parameter) with values in full agreement with those from hydrodynamical simulations (see \cite{Holanda_2017} and \cite{Zheng_2019}). Finally, for the parameter $K(z)$,  we have used the value reported by \cite{Applegate_2016} in which {\it Chandra} hydrostatic masses to relaxed clusters were calibrated with accurate weak lensing measurements from the Weighing the Giants project. The $K(z)$ parameter was estimated to be $K=0.96 \pm 0.09 \pm 0.09$ (1$\sigma$ statistical plus systematic errors) and no significant trends with mass, redshift or the morphological indicators were verified. 

Observe that by assuming $\omega _{\text{b,}0}\equiv \Omega _{\text{b,}0}h^{2}$, we can rewrite equation~(\ref{Eq fgas}) as
\begin{equation}
f_{\text{gas}}^{\text{X-ray}}\left( z\right) = 
\frac{K \ \gamma \ \omega _{\text{b,}0}}{\Omega _{\text{m,}0}h^2} \left[\frac{H(z)D_{A}(z)}{H^{\text{fid}}(z)D^{\text{fid}}_{A}(z)} \right]^{\eta}\left[ \frac{D_{A}^{\text{fid}}\left( z\right) }{D_{A}\left(z\right) }\right] ^{\frac{3}{2}}  .
\end{equation}
Therefore, for this sample, the $\chi ^{2}$ function is given by,
\begin{equation}
\chi_{\text{GMF}}^{2}=\sum_{i=1}^{40}\frac{\left[ f_{\text{gas}}^{\text{th}}(z_i)-f_{\text{gas}, \ i}^{\text{ob}}\right] ^{2}}{\sigma _{\text{tot}, \ i}^{2}}\text{ ,}
\end{equation}
with a total uncertainty given by
\begin{equation}
\sigma _{\text{tot}, \ i}^{2} = \sigma^{2}_{f_{\text{gas}}^{\text{ob}}, \ i} + \left[f_{\text{gas}}^{\text{th}}(z_i) \right]^{2} \left\{ \left( \frac{\sigma_{K} }{K}\right)^{2} + \left( \frac{\sigma_{\gamma }}{\gamma }\right)^{2} 
+   \ln^2 \left[\frac{H(z_i)D_{A}(z_i)}{H^{\text{fid}}(z_i)D^{\text{fid}}_{A}(z_i)}\right]\sigma_{\eta}^{2} \right\} \ ,
\end{equation}
where, as presented in the previous paragraph, \{$K$,$\sigma _{K}$\}$=$\{$0.96$,$0.127$\}, \{$\gamma$,$\sigma _{\gamma }$\}$=$ \{$0.848$,$0.085$\}, and \{$\eta$,$\sigma_\eta$\} $=$\{$0.442$,$0.035$\}. In addition to the BAO and GMF measures, we will adopt a Gaussian prior such as: $\omega _{\text{b,}0}=0.0226\pm 0.00034$. This value was obtained by \cite{Cooke_2016} by using BBN + abundance of primordial deuterium. 

\section{Results and Discussions}\label{Section3}

The statistical analysis is performed by the construction of the $\chi^{2}$ function,
\begin{equation}
\chi^{2}_{\text{tot.}}(\{\alpha\})=\chi^{2}_{\text{BAO,I}}+\chi^{2}_{\text{BAO,II}}+\chi^{2}_{\text{GMF}}+ \chi^{2}_{\text{BBN}}\ .
\end{equation}
From this function, we are able to construct the likelihood distribution function, $\mathcal{L}(\{\alpha\})=Be^{-\frac{1}{2}\chi^{2}_{\text{tot.}}(\{\alpha\})}$, where $B$ is the normalization factor and $\{\alpha\}$ is the set of free parameters of the cosmological model in question, that is, 
the flat  $\Lambda$CDM and XCDM models, and the non-flat $\Lambda$CDM model.

\subsection{Flat $\Lambda$CDM model}

\begin{figure}
\centering
	\includegraphics[scale=0.65]{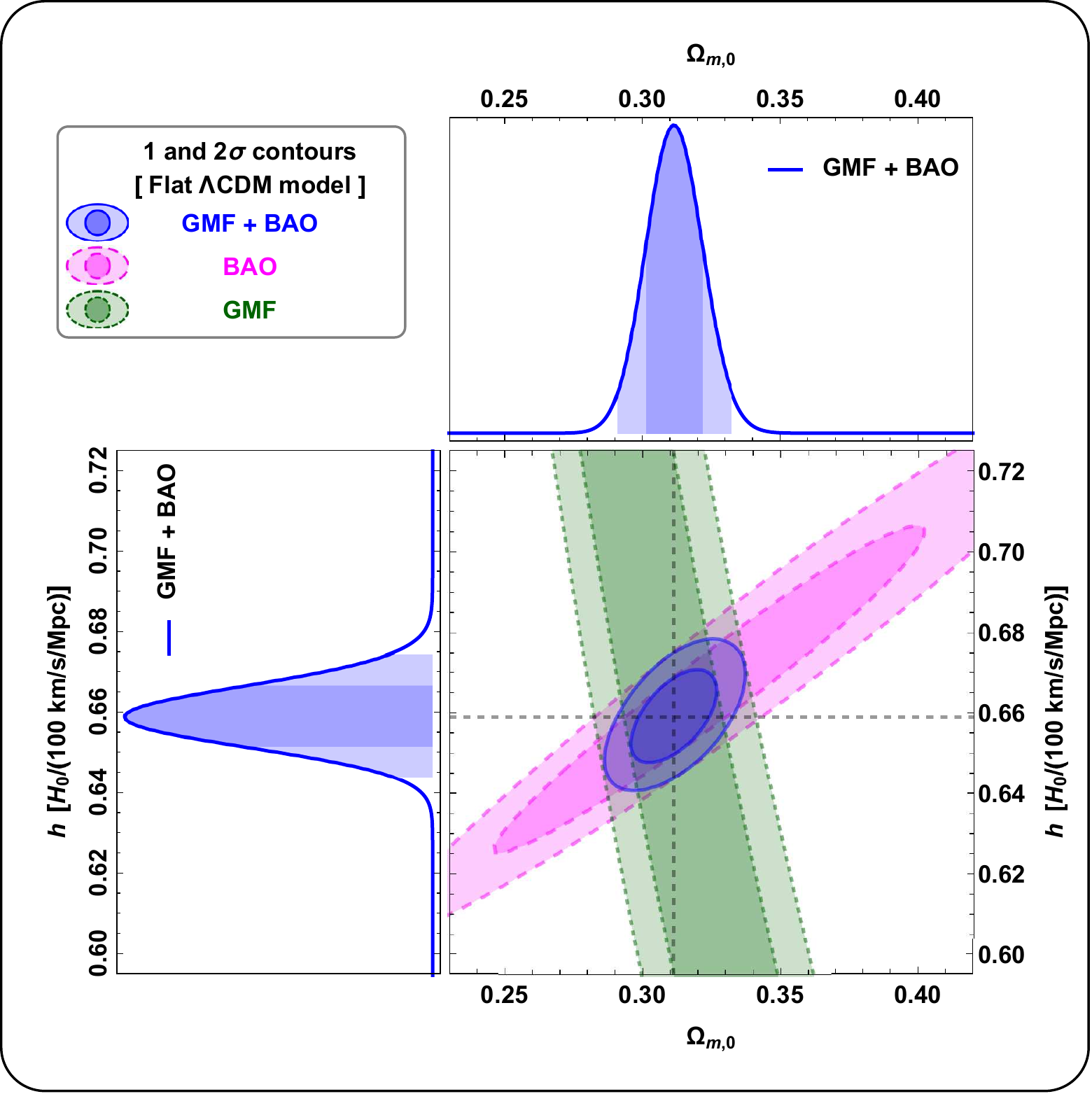}
    \caption{Contours and likelihoods of parameters $\Omega_{\text{m,}0}$ and $h$ for the flat $\Lambda$CDM model. The contours delimited by dotted green, dashed pink, and solid blue lines correspond to the analyses using only GMF, BAO and the joint analysis GMF + BAO, respectively. Regions with darker and lighter colors delimit the $1$- and $2\sigma$ c.l. regions, respectively.}
    \label{fig:LCDM}
\end{figure}

Figure~\ref{fig:LCDM} shows the contours and likelihoods for the $\Omega_{\text{m,}0}$ and $h$ parameters obtained in the context of the flat $\Lambda$CDM model. The contours delimited by dotted green lines correspond to the analysis using only GMF, the ones delimited by the dashed pink lines correspond to the analysis using only BAO, and the ones delimited by solid blue lines are referring to the joint analysis GMF + BAO. As one may see, the GMF sample alone does not restrict the value of parameter $h$ (or equivalently $H_0$) but provides tight restrictions to the value of parameter $\Omega_{\text{m,}0}$. From the joint analysis GMF + BAO, we obtain from the $\Omega_{\text{m,}0}-h$ plane (with two free parameters): $h=0.659^{+ 0.012 + 0.020}_{- 0.011 - 0.018}$  and $\Omega_{\text{m,}0} = 0.311^{+ 0.016 + 0.026}_{- 0.015 - 0.025}$ in $1\sigma$ and $2\sigma$ c.l..

By marginalizing over the parameter $\Omega_{\text{m,}0}$, we obtain the likelihood function for the $h$ parameter (see Figure~\ref{fig:h}), with: $h=0.659^{+ 0.008 + 0.015}_{- 0.007 - 0.015}$ in $1\sigma$ and $2\sigma$ c.l.. On the other hand, by marginalizing over the parameter $h$, we obtain the likelihood function of the $\Omega_{\text{m,}0}$ parameter as $\Omega_{\text{m,}0} = 0.311^{+ 0.010 + 0.021}_{- 0.010 - 0.020}$ in $1\sigma$ and $2\sigma$ c.l..

Figure~\ref{fig:h} (left) shows the likelihood of $h$ parameter for the flat (solid blue line) $\Lambda$CDM model and also the $1\sigma$ c.l. regions estimate of the $h$ parameter made by \cite{Aghanim:2018eyx} in a flat background model and \cite{Riess:2019cxk}, cosmological model independent. As one may see, our estimate is in agreement with that one from the CMB anisotropies (within $2\sigma$ c.l.)  and it is strongly discrepant with the estimate made by \cite{Riess:2019cxk}. Being more specific, our estimate of $H_0$ in a flat $\Lambda$CDM model presents a discrepancy of $5.0\sigma$ with that obtained by \cite{Riess:2019cxk}. A discrepancy also occurs  if we compare our estimate with the most recent estimate of $H_0$ obtained by SH0ES Collaboration, i.e., $H_{0}=73.5 \pm 1.4  \text{ km s}^{-1}\text{Mpc}^{-1} $ \citep{Reid_2019}. On the other hand, our estimate is in agreement with several other estimates of $H_0$ that used samples with intermediate redshifts in a flat universe \citep{Busti_2014,Abbott_2018,Yu_2018,Chen_2017,Holanda_2014,da_Silva_2018}.

\subsection{Flat XCDM model}

\begin{figure*}
\centering
	\includegraphics[scale=0.7]{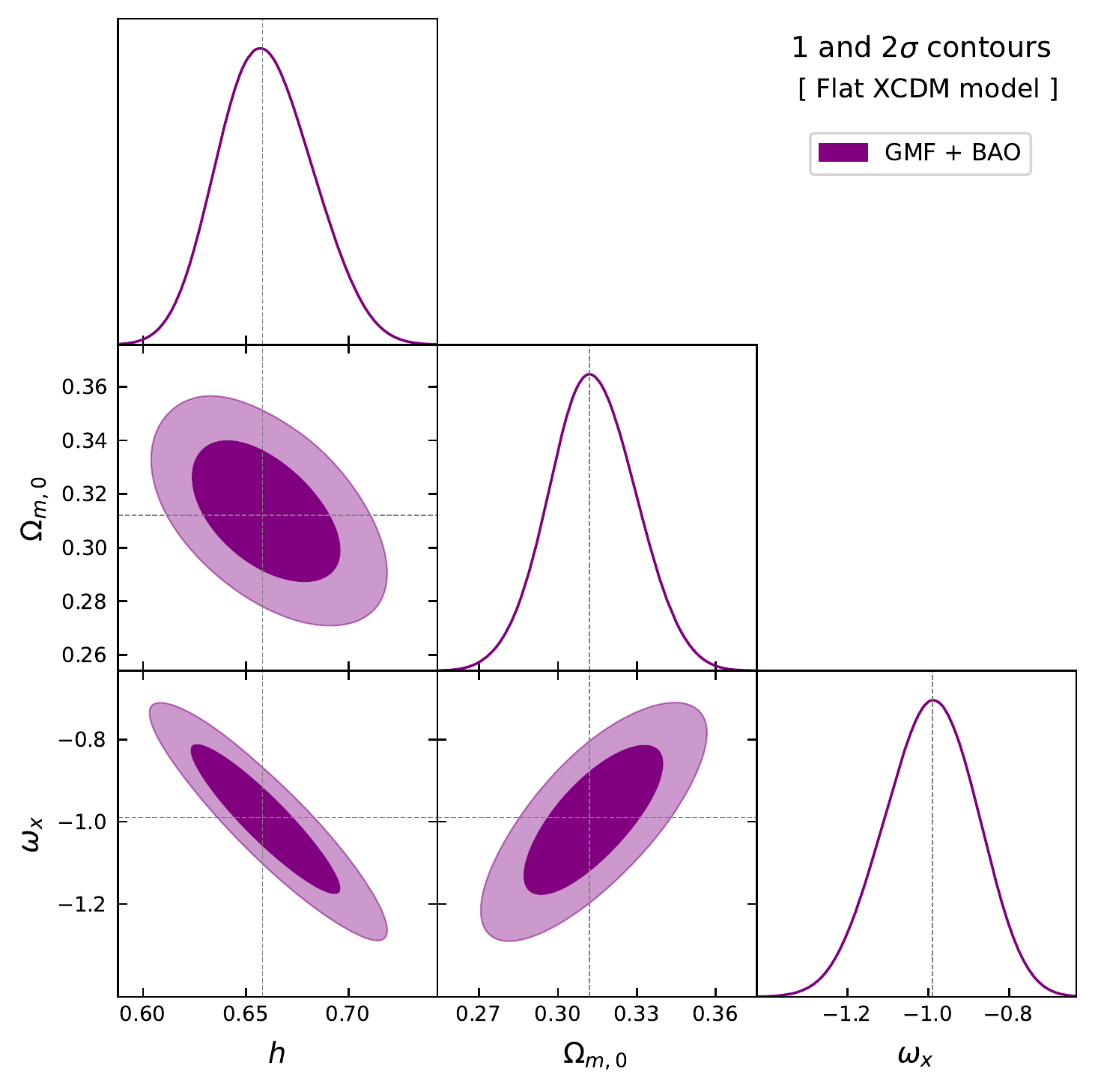}
    \caption{Marginalized contours and likelihoods of parameters $\Omega_{\text{m,}0}$, $\omega_x$ and $h$ for the flat XCDM model. The contours delimited by solid purple lines correspond to the joint analysis BAO + GMF. Regions with darker and lighter colors delimit the $1$- and $2\sigma$ c.l. regions, respectively.}
    \label{fig:XCDMflat}
\end{figure*}

Figure~\ref{fig:XCDMflat} shows the contours and likelihoods for the $\Omega_{\text{m,}0}$, $\omega_x$, and $h$ parameters obtained in the flat XCDM context. The contours delimited by solid purple lines correspond to the joint analysis BAO + GMF, where by marginalizing over the parameter $\omega_x$, we obtain from the $\Omega_{\text{m,}0}-h$ plane (with two free parameters), the intervals: $h=0.659^{+ 0.037 + 0.059}_{- 0.035 - 0.055}$, and $\Omega_{\text{m,}0} = 0.312^{+ 0.028 + 0.044}_{- 0.025 - 0.041}$  at $1\sigma$ and $2\sigma$ c.l..

On the other hand, by marginalizing over the parameter $\Omega_{\text{m,}0}$, we obtain from the $\omega_x -h$ plane (with two free parameters) the values: $h=0.659^{+ 0.037 +0.059}_{- 0.036 - 0.056}$  and $\omega_x = -0.99^{+ 0.18 + 0.28}_{- 0.18 - 0.30}$ in $1\sigma$ and $2\sigma$ c.l.. Now, by marginalizing on the parameter $h$, we obtain from the $\Omega_{\text{m,}0}-\omega_x$ plane (with two free parameters) the intervals: $\Omega_{\text{m,}0}=0.312^{+ 0.028 + 0.045}_{- 0.025 - 0.041}$ and $\omega_x = -0.99^{+ 0.18 + 0.28}_{- 0.19 - 0.30}$ in $1\sigma$ and $2\sigma$ c.l..

\begin{figure}
\centering
	\includegraphics[scale=0.75]{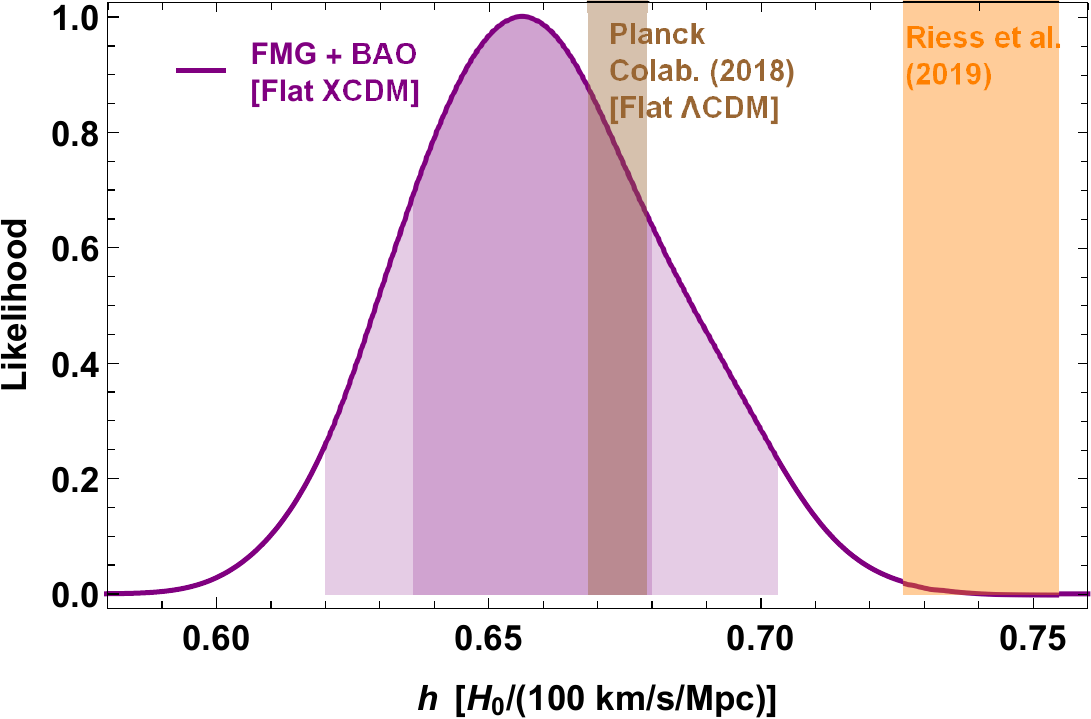}
    \caption{Marginalized likelihoods of parameter $h$ for the flat XCDM model. The regions filled with darker and lighter colors under the probability curves delimit the $1$- and $2\sigma$ c.l. regions, respectively. The brown and orange rectangle delimit the $1\sigma$ c.l. regions of the estimate of the $h$ parameter made by \cite{Aghanim:2018eyx} (flat $\Lambda$CDM: $h=0.6736\pm 0.0054$) and \cite{Riess:2019cxk} (independent-model: $h=0.7403\pm 0.0142$), respectively.
    }
    \label{fig:h2}
\end{figure}

In order to obtain the likelihood function for the $h$ parameter, we marginalize over $\Omega_{\text{m,}0}$ and $\omega_x$ parameters (see Figure~\ref{fig:h2}). From this  likelihood, the following estimate is found: $h=0.659^{+ 0.021 + 0.044}_{- 0.024 - 0.040}$ in $1\sigma$ and $2\sigma$ c.l.. Our estimate is in agreement with that one from the CMB anisotropies (within $2\sigma$ c.l.)  and it is  discrepant with the estimate made by \cite{Riess:2019cxk}. Actually, our estimate of $H_0$ in a flat XCDM model presents a discrepancy of $3.1\sigma$ with that obtained by \cite{Riess:2019cxk}.

Similarly, by marginalizing over the $h$ and $\omega_x$ parameters, we obtain the likelihood for the parameter $\Omega_{\text{m,}0}$ with the following intervals: $\Omega_{\text{m,}0} = 0.312^{+ 0.016 + 0.032}_{- 0.016 - 0.032}$ at $1\sigma$ and $2\sigma$ c.l.. Finally, by marginalizing over the $h$ and $\Omega_{\text{m,}0}$ parameters, we obtain the likelihood for the parameter $\omega_x$ as $\omega_x = -0.99^{+ 0.11 + 0.20}_{- 0.11 - 0.21}$ in $1\sigma$ and $2\sigma$ c.l., in full agreement with the flat $\Lambda$CDM model ($\omega_x=-1$).

\subsection{Non-flat $\Lambda $CDM model}

\begin{figure*}
\centering
	\includegraphics[scale=0.5]{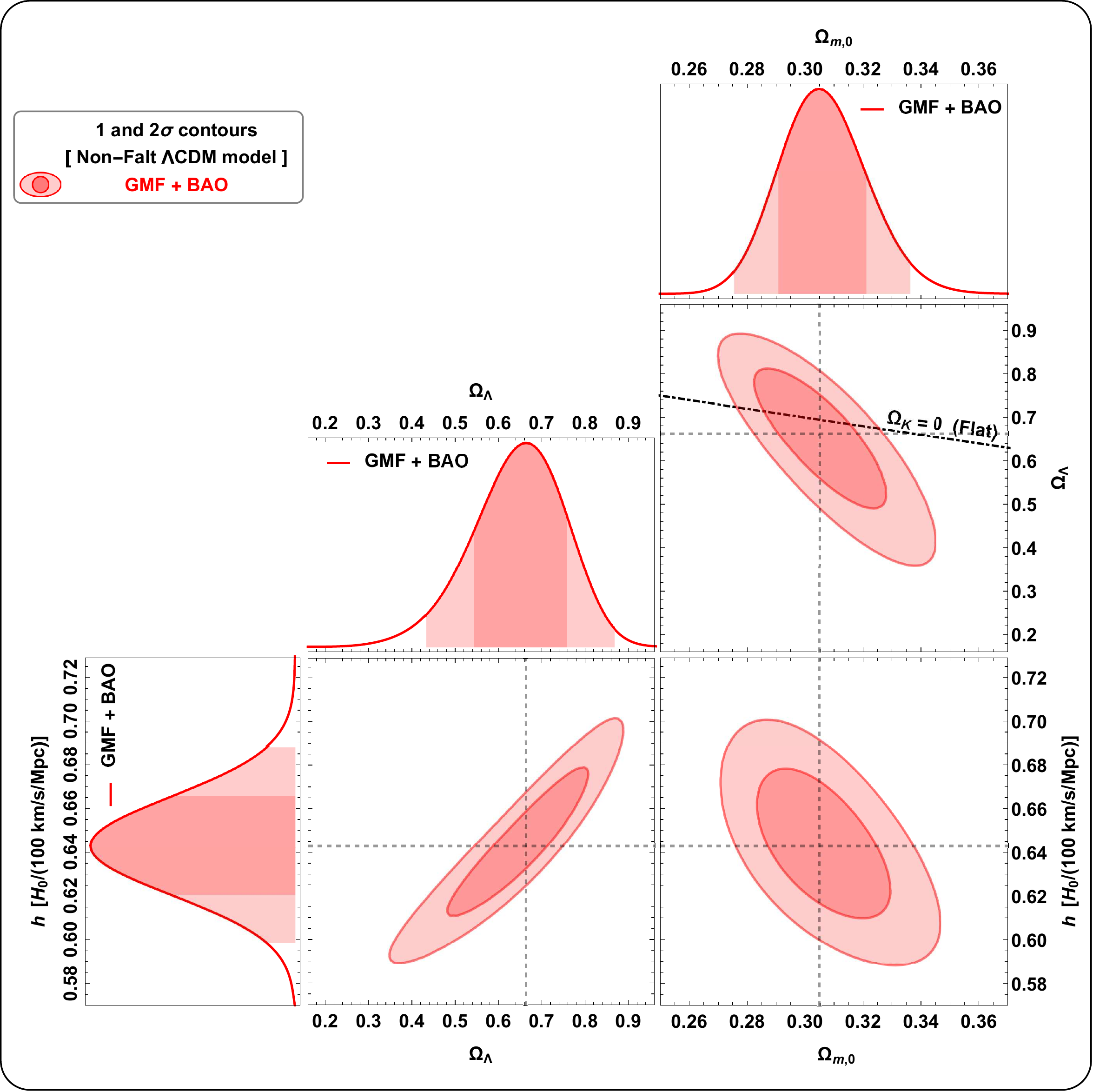}
    \caption{Marginalized contours and likelihoods of parameters $\Omega_{\text{m,}0}$, $\Omega_{\Lambda}$ and $h$ for the non-flat $\Lambda$CDM model. The contours delimited by solid red lines correspond to the joint analysis BAO + GMF. Regions with darker and lighter colors delimit the $1$- and $2\sigma$ c.l. contours, respectively.}
    \label{fig:NonFlat}
\end{figure*}

Figure~\ref{fig:NonFlat} shows the contours and likelihoods for the $\Omega_{\text{m,}0}$, $\Omega_{\Lambda}$, and $h$ parameters obtained in the context of the non-flat $\Lambda$CDM model. Similar to the case of the previous section, the contours delimited by solid red lines correspond to the joint analysis BAO + GMF. For this analysis, by marginalizing over the parameter $\Omega_{\Lambda}$, we obtain from the $\Omega_{\text{m,}0}-h$ plane (with two free parameters), the intervals: $h=0.644^{+ 0.035 + 0.057}_{- 0.034 - 0.056}$, and $\Omega_{\text{m,}0} = 0.305^{+ 0.024 + 0.042}_{- 0.022 - 0.034}$  in $1\sigma$ and $2\sigma$ c.l..

Similarly, by marginalizing over the parameter $\Omega_{\text{m,}0}$, we obtain from the $\Omega_{\Lambda}-h$ plane (with two free parameters) the values: $h=0.645^{+ 0.034 +0.057}_{- 0.034 - 0.056}$  and $\Omega_{\Lambda} = 0.660^{+ 0.146 + 0.227}_{- 0.179 - 0.312}$ in $1\sigma$ and $2\sigma$ c.l..

Finally, by marginalizing on the parameter $h$, we obtain from the $\Omega_{\text{m,}0}-\Omega_{\Lambda}$ plane (with two free parameters) the intervals: $\Omega_{\text{m,}0}=0.305^{+ 0.023 + 0.040}_{- 0.022 - 0.035}$ and $\Omega_{\Lambda} = 0.661^{+ 0.152 + 0.232}_{- 0.170 - 0.301}$ in $1\sigma$ and $2\sigma$ c.l..

\begin{figure*}
	\includegraphics[scale=0.68]{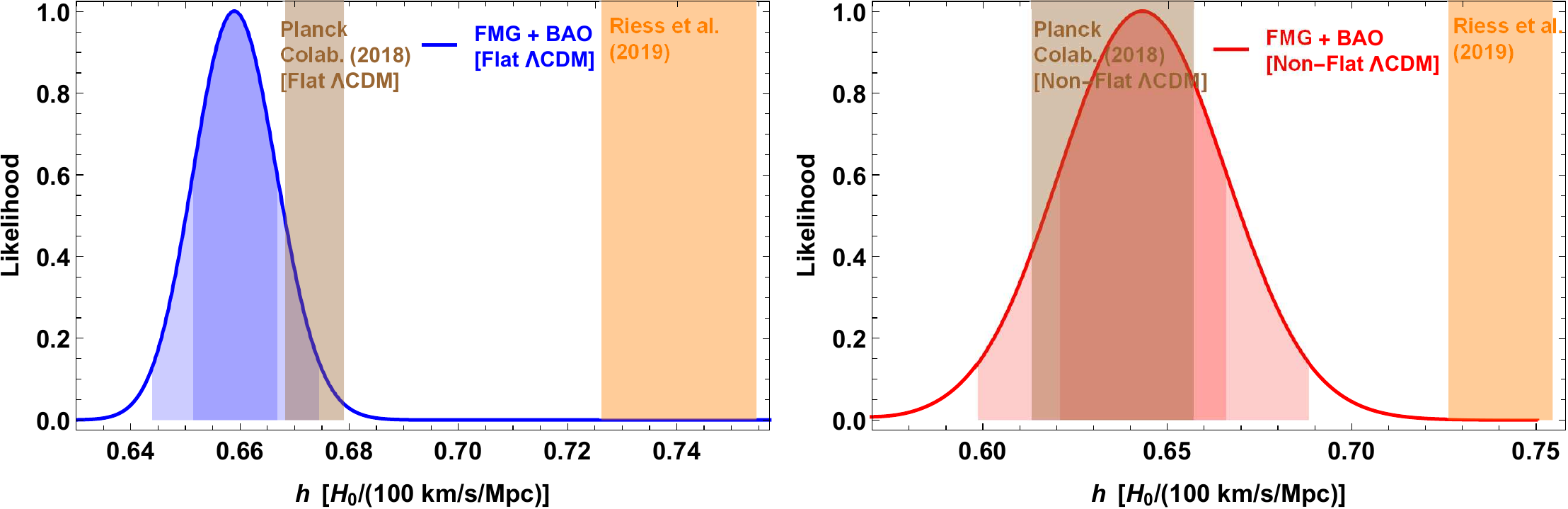}
    \caption{Marginalized likelihoods of parameter $h$ for the flat (left) and non-flat (Right) $\Lambda$CDM model. The regions filled with darker and lighter colors under the probability curves delimit the $1$- and $2\sigma$ c.l. regions, respectively. The brown and orange rectangles correspond to  $1\sigma$ c.l. regions of the $h$ estimates  made by \cite{Aghanim:2018eyx} (flat $\Lambda$CDM: $h=0.6736\pm 0.0054$; non-flat $\Lambda$CDM: $h=0.636^{+0.021}_{-0.023}$) and by \cite{Riess:2019cxk} (independent-model: $h=0.7403\pm 0.0142$), respectively.
    }
    \label{fig:h}
\end{figure*}

On the other hand, in order to obtain the likelihood function for the $h$ parameter, we marginalize over $\Omega_{\text{m,}0}$ and $\Omega_{\Lambda}$ parameters (see Figure~\ref{fig:h}). From this  likelihood, the following estimate is found: $h=0.643^{+ 0.023 + 0.045}_{- 0.022 - 0.045}$
in $1\sigma$ and $2\sigma$ c.l.. Similarly, by marginalizing over the $h$ and $\Omega_{\Lambda}$ parameters, we obtain the likelihood for the parameter $\Omega_{\text{m,}0}$ with the following intervals: $\Omega_{\text{m,}0} = 0.305^{+ 0.016 + 0.031}_{- 0.014 - 0.029}$ in $1\sigma$ and $2\sigma$ c.l.. Finally, by marginalizing over the $h$ and $\Omega_{\text{m,}0}$ parameters, we obtain the likelihood for the parameter $\Omega_{\Lambda}$ as $\Omega_{\Lambda} = 0.663^{+ 0.094 + 0.204}_{- 0.120 - 0.230}$ in $1\sigma$ and $2\sigma$ c.l..

Figure~\ref{fig:h} (right) shows the likelihood of $h$ parameter for the non-flat (dashed red line) $\Lambda$CDM model, together with $1\sigma$ c.l. regions of the estimate of the $h$ parameter obtained from CMB anisotropies \cite{Aghanim:2018eyx} in a non-flat background and that one from  the Ref.~\cite{Riess:2019cxk} (local method). From the figure, it is evident that our estimate is in full agreement (within $1\sigma$ c.l.) with that one  from the CMB anisotropies \cite{Aghanim:2018eyx} and discrepant with the local estimate made by \cite{Riess:2019cxk} in $3.6\sigma$ c.l..

The Table~\ref{Tab:parametros} shows a synthesis of the results presented in the last three subsections. More specifically, we show the free parameter estimates of all cosmological models in $2\sigma$ c.l. obtained from their respective likelihoods. As one may see, the $H_0$ values in all cases  analyzed  are compatible within 2$\sigma$ with the  Hubble constant measurement from the Planck results, presenting  tensions with the local estimate. The estimates obtained here are considerably tighter than those ones from the Ref.~\cite{Holanda_2012}, where angular diameter distances of galaxy clusters plus BAO and shift parameter were used. The $\Omega_{\text{m},0}$ parameter for the flat model is also compatible with the estimate $\Omega_{\text{m,}0} = 0.3158\pm 0.0073$ from \cite{Aghanim:2018eyx} within $1\sigma$ c.l.. Also for the non-flat case, the estimates for $\Omega_{\text{m,}0}$ and $\Omega_{\Lambda}$ are in  agreement to \cite{Aghanim:2018eyx} in $2\sigma$ c.l.. 

It is important to comment that we also perform our analyses by using another Gaussian prior, such as,  $\omega _{\text{b,}0}=0.02156\pm 0.0002$ \cite{Cooke_2016}. Although this value differs by $\approx 2.3\sigma$ from the Standard Model value estimated from the Planck observations of the cosmic microwave background, our $H_0$ estimates had insignificant changes.

\subsection{Deceleration parameter and curvature density parameter}

Using the  uncertainties propagation and the values estimated for $\Omega_{\text{m,}0}$, $\omega_x$ and  $\Omega_{\Lambda}$ from their likelihoods, we can estimate the current value of the deceleration parameter for each model analyzed by using $q_0 = \Omega_{\text{m,}0}/2 +(1-3\omega_a)\Omega_{a}/2$, where \{$\omega_a \rightarrow -1$, $ \Omega_{a} \rightarrow 1- \Omega_{\text{m,}0}$ \}, \{$\omega_a \rightarrow \omega_{x}$, $ \Omega_{a} \rightarrow 1- \Omega_{\text{m,}0}$\} and \{$\omega_a \rightarrow -1$, $ \Omega_{a} \rightarrow \Omega_{\Lambda}$ \} for the flat $\Lambda$CDM and XCDM models, and the non-flat $\Lambda$CDM model, respectively. We obtain $q_0 = -0.533 \pm 0.046$, $q_0 = -0.542 \pm 0.299$ and $q_0 = -0.483 \pm 0.340 $ in $3\sigma$ c.l. for the flat $\Lambda$CDM and XCDM models, and non-flat $\Lambda$CDM model, respectively. Moreover, similarly, we estimate the following value for the current curvature density parameter in $1\sigma$ c.l.: $\Omega_{\text{k},0}=0.056 \pm 0.108$. Our  results from these models indicate an accelerating expansion of the universe in more than $3\sigma$ c.l., and $\Omega_{\text{k,}0}$ compatible with a spatially flat curvature within $1\sigma$ c.l..

\begin{table}
\caption{A summary of the constraints in $2\sigma$ c.l. on the set of free parameters of the flat and non-flat $\Lambda$CDM model.}
\label{Tab:parametros}
\def\arraystretch{1.2} 
\centering
\begin{tabular}{cccc}
\hline \hline
                         &  Flat $\Lambda$CDM model      &  Non-flat $\Lambda$CDM model &  Flat XCDM model  \\ \hline
$H_{0}$ $\left[\text{km s}^{-1}\text{Mpc}^{-1} \right]$ & $65.9^{+ 1.5}_{- 1.5}$     &  $64.3^{+4.5}_{-4.4}$ &  $65.9^{+4.4}_{-4.0}$   \\ \hline
 $\Omega_{\text{m,}0}$   &$0.311^{+ 0.021}_{- 0.020}$ &$0.305^{+ 0.031}_{- 0.029}$ &$0.312^{+ 0.032}_{- 0.032}$ \\ \hline
                $\Omega_{\Lambda}$      &  ---                          &  $0.663^{+ 0.204}_{- 0.230}$&  ---                      \\ \hline
   $\omega_x$      &  ---                          &   ---      &  $-0.99^{+0.20}_{-0.21}$                 \\ \hline
 $\chi^{2}_{\text{min}} $&  $25.91$                      & $ 25.90$     & $ 25.90$                    \\ \hline 
\hline    
\end{tabular}
\end{table}

\section{Conclusions}\label{Section4}

The recent $H_0$ tension between the local and global methods for the Hubble constant  value has motivated the search for new tests and data analyses that could alleviate (or solve) the discrepancy in the $H_0$ value estimated by different methods.

In this paper, we  obtained new and tight estimates on the Hubble constant by combining 40 galaxy cluster gas mass fraction measurements with 11 baryon acoustic oscillation data in the frameworks of the flat XCDM and $\Lambda$CDM models, and the non-flat $\Lambda$CDM models. The data sets are in the following range of redshift $0.078\leq z \leq 1.023 $. For all cosmological models considered, the gas mass fraction sample alone did not restrict the value of $H_0$, but put restrictive limits on the $\Omega_M$ parameter. However, from the joint analysis with the 11 BAO data, the restriction on the possible $H_0$ values was notable. By considering the flat $\Lambda$CDM and XCDM models, and the non-flat $\Lambda$CDM model, we obtained, respectively: $H_0=65.9^{+1.5}_{-1.5}$ km s$^{-1}$ Mpc$^{-1}$, $H_0=65.9^{+4.4}_{-4.0}$ km s$^{-1}$ Mpc$^{-1}$ and $H_0=64.3^{+ 4.5}_{- 4.4}$ km s$^{-1}$ Mpc$^{-1}$  in $2\sigma$ c.l.. For all cases, the estimates indicated a low value of $H_0$ as those ones obtained by the Planck satellite results from the CMB anisotropies observations. For the flat models, the agreement is within $2\sigma$ c.l., while for the non-flat model the concordance is  within $1\sigma$ c.l. (see Figure~\ref{fig:h}). In this point, it is worth to point that in our analyses priors from BBN to calibrate the cosmic sound horizon and the  cosmic microwave background radiation local  temperature as given by the COBE/FIRES also were used. Our results reinforce the $H_0$ tension regardless the curvature cosmic and  the  equation of state parameter of dark energy. 
 
Estimates for the current deceleration and curvature density parameters  were also obtained. The results from the considered models pointed to a Universe in accelerated expansion in more than $3\sigma$ c.l.. We obtained  $q_0 = -0.533 \pm 0.046$ and $q_0 = -0.542 \pm 0.299$ for the flat $\Lambda$CDM and XCDM models in $3\sigma$ c.l., respectively. For the non-flat $\Lambda$CDM model it was estimated $q_0 = -0.483 \pm 0.340 $ in $3\sigma$ c.l.. Moreover, for the non-flat model, although the best fit suggested a positive curvature, our analysis is compatible with a spatially flat curvature within $1\sigma$ c.l..

Finally, it is important to stress that the $H_0$ estimates obtained here considered no  evolution of  gas mass fraction measurements within redshift and mass intervals  of the galaxy cluster sample used. This question is still open for some debate.  In the coming years, the eROSITA \citep{merloni2012erosita} mission will make an all-sky X-ray mapping of thousand of galaxy clusters and will provide accurate information on gas mass fraction measurements, which will turn the analysis proposed here more robust.

\acknowledgments

RFLH thanks financial support from Conselho Nacional
de Desenvolvimento Cientıfico e Tecnologico (CNPq) (No.428755/2018-6 and 305930/2017-6). SHP would like to thank CNPq for financial support, No.303583/2018-5. The figures in this work  were created with Wolfram Mathematica and GetDist \citep{Lewis:2019xzd}.
\bibliography{RefBibTex}

\end{document}